# A tool for user friendly, cloud based, whole slide image segmentation

*Brendon Lutnick[1], David Manthey[2], and Pinaki Sarder[1,*]*

[1]Department of Pathology and Anatomical Sciences, SUNY Buffalo,
[2]Kitware Incorporated, Clifton Park, New York.

[*]Address all correspondence to: Pinaki Sarder
Tel: 716-829-2265; E-mail: pinakisa@buffalo.edu

## ABSTRACT

Convolutional neural networks, the state of the art for image segmentation, have been successfully applied to histology images by many computational researchers. However, the translatability of this technology to clinicians and biological researchers is limited due to the complex and undeveloped user interface of the code, as well as the extensive computer setup required. We have developed a tool for segmentation of whole slide images (WSIs) with an easy to use graphical user interface. Our tool runs a state-of-the-art convolutional neural network for segmentation of WSIs in the cloud. Our plugin is built on the open source tool HistomicsTK by Kitware Inc. (Clifton Park, NY), which provides remote data management and viewing abilities for WSI datasets. The ability to access this tool over the internet will facilitate widespread use by computational non-experts. Users can easily upload slides to a server where our plugin is installed and perform human in the loop segmentation analysis remotely. This tool is open source, and has the ability to be adapted to segment of any pathological structure. For a proof of concept, we have trained it to segment glomeruli from renal tissue images, achieving an *F-score > 0.97* on holdout tissue slides.

## INTRODUCTION

Recent advancements in machine learning techniques have attained previously unachievable accuracy for image analysis tasks. In particular, convolutional neural networks (CNNs)[1] - a form of deep learning[2], have great potential for impactful applications for the segmentation of image structures. In the field of pathology, CNNs have been successfully utilized by many research groups for the segmentation of whole slide images (WSIs)[3-6]. However, thus far tools to segment WSIs have been complex to deploy and use, requiring use of the command line interface and computational expertise[7-9]. Going beyond development, the target demographic for these tools is the pathologist or biological scientist, whose clinical workflow or research questions could leverage fast and accurate segmentation of relevant structures. To address this gap, we have developed a powerful tool for the segmentation of WSIs and deployed it as an easy to use plugin in a cloud based WSI viewer. Upon the completion of our code, it will be open-sourced and easily deployable on a remote server for use by the community over the web. Additionally, we will host an instance of our tool which is publicly available for the community, for the sake of security, this instance will only be accessible after approval of a user's account.

This tool is an extension our out previous work H-AI-L[6], where we showed that iterative annotation of WSIs significantly reduced the annotation burden. Like most works in computational digital pathology, H-AI-L found limited use in the community due to the complexities of installation. To address this, our new segmentation tool does not require the installation of any software on the user's local computer, and all the processing is handled on the remote server which is hosting the web client. It produces computational annotations which are automatically displayed on top of the slide using the HistomicsTK interface[10] as seen in. The user can pan and zoom the slide as well as interact with the annotations, removing, correcting, or adding regions. See Figs. 1 and 2 for details about this graphical user interface.

While our segmentation plugin is agnostic to tissue type or structure of interest, we have validated it by training a CNN model for the segmentation of glomeruli from renal tissue images. However, in an effort to make this tool more useful to the digital pathology community, we have also created a plugin for training new segmentation models. Using our simple cloud-based interface users can upload and annotate WSIs, and train a segmentation network using their annotations, see

Fig. 2. Like in H-AI-L[6], users can iteratively use the training and prediction plugins of our segmentation tool in an active learning framework, to build up powerful segmentation models with minimal effort.

## RESULTS & DISCUSSION

To access the potential of our segmentation tool we used a network model for glomeruli segmentation (trained using 768 WSIs) to segment 100 holdout WSIs of diverse stain, institution, scanner, and species. The 100 holdout slides included 3816 glomeruli, 37.8 GB of compressed image data, and a combined total of more than 0.24 trillion image pixels. We compared the computationally generated annotations with hand annotations for glomeruli and observed the following performance (Fig. 3B):

*F-score=0.97 | MCC=0.97 | Kappa=0.97 | IOU=0.941 | Sensitivity=0.953 | Specificity=1.0 | Precision=0.988 | Accuracy=1.0*

To our knowledge, our study of glomeruli segmentation not only uses the largest most diverse cohort of slides, but also reports the best performance of any study reported in the literature. In our previous work H-AI-L[6] we trained Deeplab-v2[11] using a dataset of 13 PAS and H&E stained mouse WSIs containing 913 glomeruli, and achieved an F-score=0.92. *Kannan et al.*[12] (who used Inception-V3[13] for the sliding window classification of glomeruli with a training set of 885 patches from 275 trichrome stained biopsy's) report MCC = 0.628. *Bueno et al.*[14] trained U-net[5] with 47 PAS stained WSIs reported accuracy=0.98. *Gadermayr et al.*[15] used 24 PAS stained mouse WSIs to train U-net[5], reporting precision=0.97 and sensitivity=0.86.

Of all the works in the literature, *Jayapandian el al.*[16] present the most comprehensive results on glomerular segmentation. They trained U-net[5] on a dataset containing 1196 glomeruli from 459 human WSIs stained with H&E, PAS, Silver and Trichrome, reporting an F-score of 0.94. However their analysis is limited to data with minimal change disease[17], which as the name describes, presents pathologically as normal glomeruli. In contrast our training dataset contained 768 WSIs, stained with H&E, PAS, Silver, Trichrome, Toluene Blue, CD-68, Verhoeff's Van Gieson, Jones, and Congo red. In total this dataset contains 61734 glomeruli, from nearly 50 disease pathologies with both human and mouse data. Our holdout dataset was split at the slide level, and contains mouse and human data from different institutions, scanners, and stains, with multiple disease pathologies present. Examples of holdout glomeruli are shown in Fig. 4.

Our segmentation tool works natively on WSIs without the need for patch extraction prior to training / prediction. It produces segmentations which contain the contours of detected regions for the whole WSI. When developing new segmentation models, the slide-viewing environment of our tool, enables rapid qualitative evaluation of algorithm progress by displaying network predictions directly on the slides as a series of annotation contours (Fig. 1). With the ability to correct computationally generated annotations on holdout slides, it is easy to add corrected slides to the training dataset (Fig. 2). Our tool uses hardware acceleration on the host server to speed up processing, and is capable of segmenting large histology slides in as little as 1 min. The segmentation time depends (roughly linearly) on the size of the tissue section in the slide, Fig. 3A quantifies the detection speed as a function of image pixels on a large cohort of 1591 WSIs. Our algorithm performs a fast thresholding of the tissue region contained within the slide to reduce the computational burden on slides with large non-tissue areas, there is a slight programmatical computational overhead when opening and caching larger slides – seen as gentle upslope of points of the same color in Fig 3A.

We have found that using this tool to alternate iteratively between training and prediction greatly reduces the annotation burden, allowing experts to correct the network predictions on holdout WSIs before incorporating them into the training-set[6]. When selecting new data to add to the training-set, we found the ability to view predictions interactively on the WSI is extremely helpful to determine slides where the current model struggles. Practically, we have found that the performance characteristics of our tool (a very high specificity in glomerular segmentation – Fig. 3) are very favorable when deployed in a human in the loop setting. This enables the user to add missing annotations, without needing to remove many false positive predictions. Indeed, when building up our glomerular training-set, we found that the network quickly began to identify whole glomeruli, missing those with severe disease pathology and abnormal staining.

Annotations done directly on the WSI in an interactive viewing environment easily fits into pathologist workflow, and the cloud-based nature of our tool abstracts any computational overhead away from the end user. Annotation can be done on **any** internet connected device without any software installation - including a mobile phone. If the user prefers to annotate locally, we have added options to ingest and export annotations in a XML[18] format readable by the commonly used WSI viewer Aperio Imagescope[19]. The authors note a complimentary work: Quick Annotator[20], a recently published work which speeds the annotation of histology slides using the locally installed QuPath slide viewer[21]. This tool uses superpixels[22] and deep learning to speed the segmentation of local regions of interest within a WSI. In the future we would like to utilize a similar approach combined with edge detection and snapping[23] to speed the initial segmentation by human annotators. We also note that at the time of writing this preprint we are also working to incorporate our core segmentation code into QuPath.

A video overview of a beta version of the tool available here:
*https://buffalo.box.com/s/w9ao5p1qs9o3lgyqk8ioih6grklb9p9i*.

# METHODS

With the goal of developing a tool with class leading WSI segmentation accuracy as well as easy accessibility to computational non-experts, we have integrated the popular semantic segmentation network Deeplab V3+[24] with Digital Slide Archive[10] the open-source cloud-based histology management program. Specifically, we have created an easy to use plugin using HistomicsTK, an application programing interface of the Digital Slide Archive for running Python codes. This plugin efficiently runs the Deeplab network for native segmentation of WSIs, making testing new slides accessible through the HistomicsTK graphical user interface.

*Software:* This plugin, built using open-source software, is available to the digital pathology community for use and further development. When the final manuscript on this work is publishes, our modified HistomicsTK-Deeplab codebase will be made available on GitHub and also as a prebuilt Docker image for easy installation. This software is deployed in the cloud and is accessible via the web, making it easily accessible to the community (Fig. 1).

*Functionality:* Our plugin outputs predictions as a series of image contours which we format to a JSON[25] format for display in HistomicsTK. We have made our code modular with the ability to handle multi-class segmentation, and included optional prediction parameters for advanced users. We have also included functionality for conversion to the XML format used to display contours in Aperio ImageScope, a popular WSI viewer, as well as export annotated regions as image masks.

*Dataset:* The glomeruli model presented here was trained using 768 WSIs containing 61734 annotated glomeruli, specifically the training set contained 428 human, and 315 mouse WSIs with various disease pathologies, and histologic stains. To evaluate this model we used a separate set of 100 randomly selected holdout WSIs containing 3816 annotated glomeruli.

*Computational model:* The Deeplab V3+ network was trained for 400000 steps using cropped patches of size 512×512 pixels, a batch size of 12, and learning rate of $1e^{-3}$. This model used the Xception network backbone with an output stride of 16 as described in the Deeplab paper[24].

*Human data:* Biopsy samples from human diabetic nephropathy patients were obtained from 7 institutions. Human data collection procedure followed a protocol approved by the Institutional Review Board at University at Buffalo. Ground-truth annotations of glomeruli were preformed using the annotation tools provided by HistomicsTK and Aperio ImageScope, where computational predictions were corrected as described in our previous work[6].

*Imaging and data preparation:* Human tissue slices of 2-5 μm were stained using diverse histological stains, and imaged using a whole-slide imaging scanner. When possible, we followed similar imaging protocol as described in our previous work[26].

# CONCLUSION AND FUTURE WORK

We have developed a user-friendly tool for cloud-based segmentation of WSI. To show the utility of this method, we trained a model for the segmentation of glomeruli from renal tissue slides which we demonstrate and quantify on holdout slides. At the time of writing, we have deployed this tool on two servers: one rented from the SUNY Buffalo center for computational research, and another owned by our team. A demonstration is available in the video linked in the results section. We plan to opensource the code and open this server to the public upon publishing the final manuscript on this work. This plugin is open source and packaged as a pre-built Docker image which is easy to setup on a server. The software is plug-and-play, allowing users to upload data in the Digital Slide Archive, run segmentation via HistomicsTK, and correct segmentation error via HistomicsTK web-viewer. We designed the plugin with flexibility in mind, it was designed to efficiently run segmentation on WSIs of any tissue type. Segmentation of new structures of interest is possible by retraining the Deeplab network used for segmentation, which can be conveniently done within the HistomicsTK interface. We are currently working to train additional models for segmentation of Interstitial fibrosis and tubular atrophy (IFTA), as well as arteries and arterioles in renal tissue, which we will use to study the usability of our tool by expert pathologists.

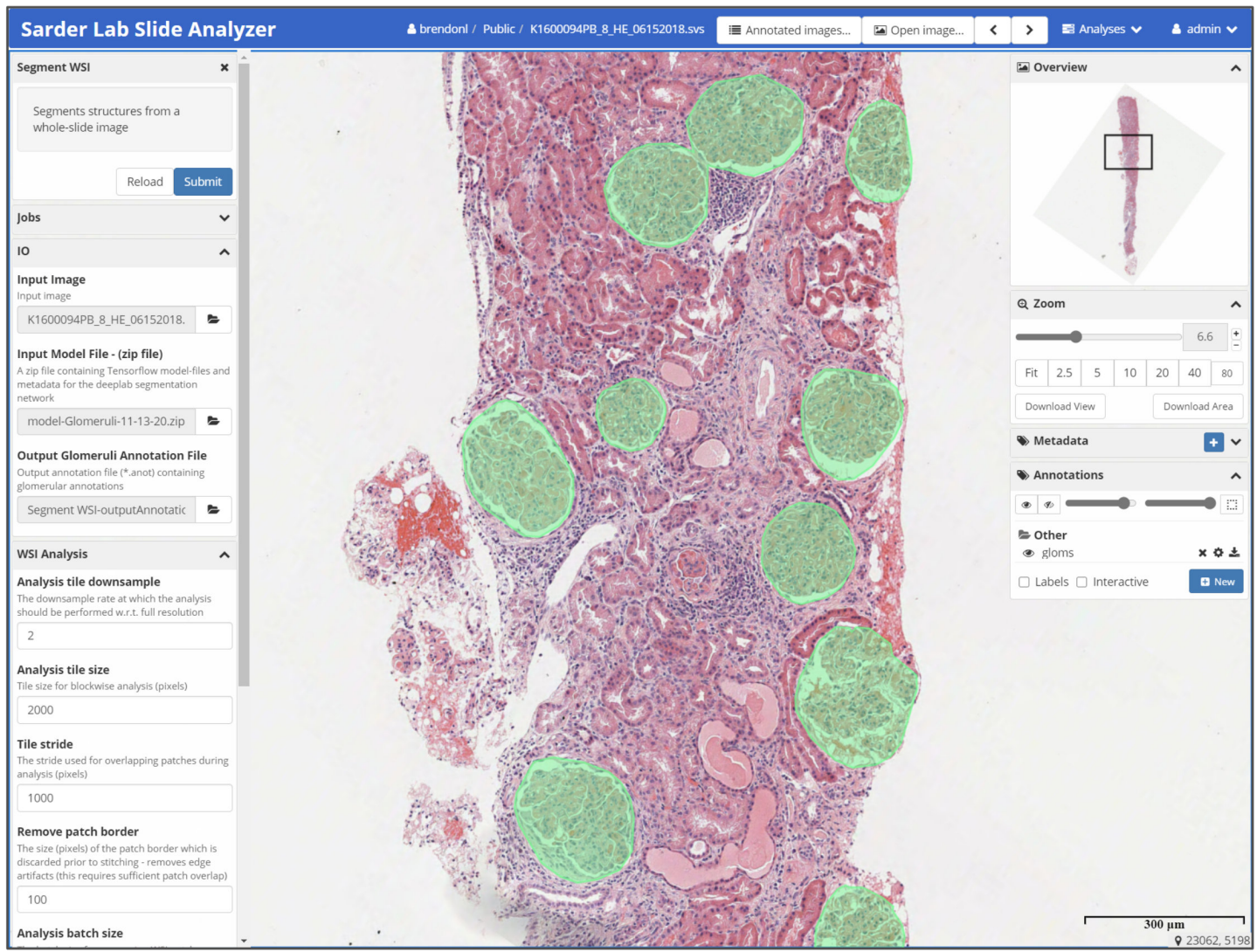

***Fig. 1 |*** The user interface of our segmentation plugin (available via the web).
The left *Segment WSI* column shows the controls for the plugin: *IO* is required arguments and *WSI Analysis* contains optional parameters. The right column shows the WSI viewer controls and annotations created by our plugin. The green annotations on the holdout slide are predicted by our plugin and are easily editable by the user. Slides are analyzed by clicking the *Submit* button in the top left corner, which submits a segmentation job, running the Deeplab network on the remote server (where HistomicsTK is installed).

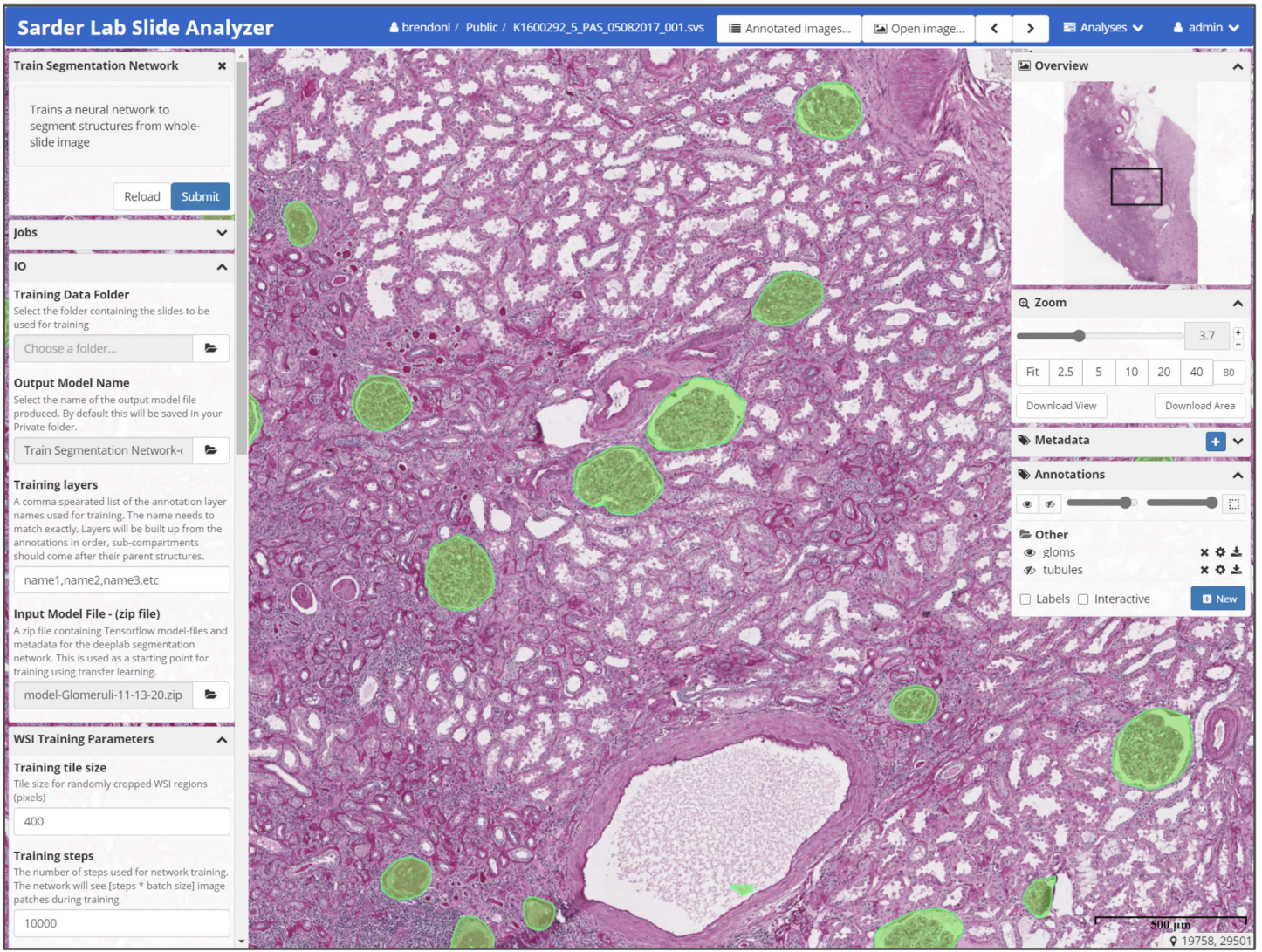

***Fig. 2 |*** The interface of the training plugin.
The left panel shows the *Train Segmentation Network* plugin. Under the *IO* section, a user can specify a directory full of annotated WSIs to use for network training with the *Training Data Folder* option, and where to save the trained model with the *Output Model Name* option. The *Training layers* option gives users the ability to choose which of the annotation layers should be used for training, single or multi-class segmentation models can be trained. To speed up the training process, a previously trained segmentation model can be used for transfer learning by specifying the *Input Model File*. Hyper-parameters for training the network can be specified using the options in the *WSI Training Parameters* section – these come set to defaults which we have found work well.

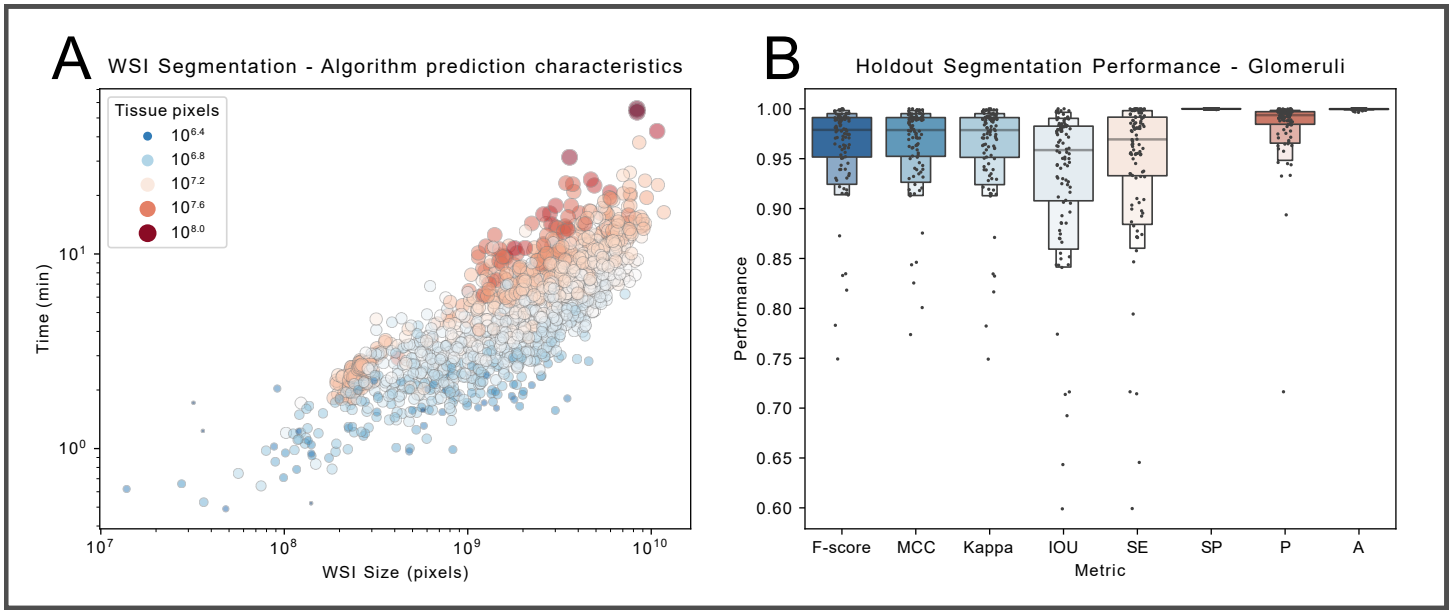

**Fig. 3 |** Performance characteristics of the segmentation plugin.
A shows the prediction time in minutes as a function of the WSI size in pixels for glomerular predictions on 1528 WSIs. The color and size of the points represent the size of the automatically extracted tissue region of the slide (the analyzed region) in pixels. Our plugin scales roughly linearly in time for increasing WSI size. B the segmentation performance of our glomerular detection model when applied to 100 holdout WSIs. Here MCC, IOU, SE, SP, P, and A represent Matthews Correlation Coefficient, Intersection-Over-Union, Sensitivity, Specificity, Precision, and Accuracy metrics respectively. We also calculated the summed performance of our method by summarizing the true/false - positive/negative predictions across all 100 holdout slides: *F-score=0.970 | MCC=0.970 | Kappa=0.970 | IOU=0.941 | Sensitivity=0.953 | Specificity=1.000 | Precision=0.988 | Accuracy=1.000*

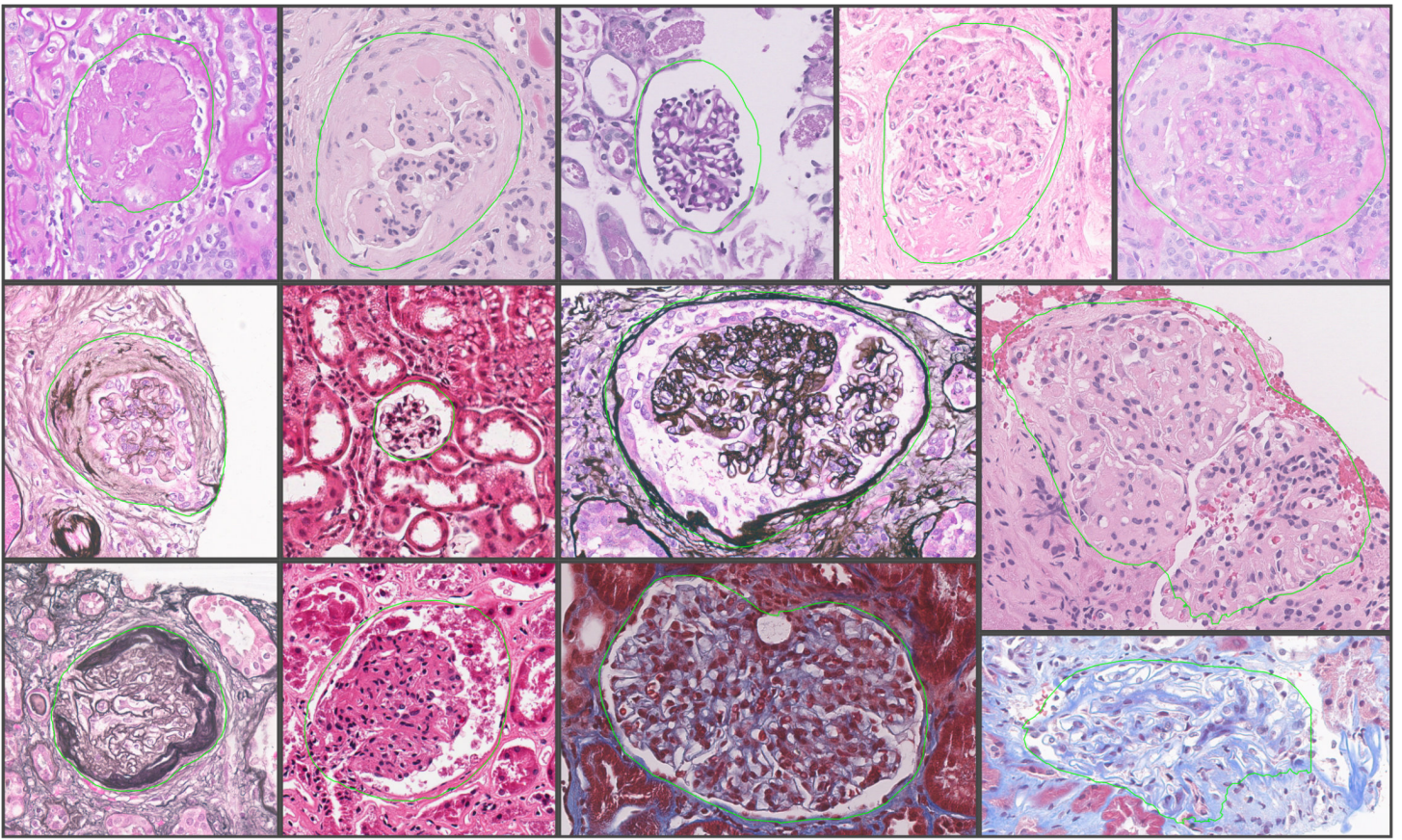

**Fig. 4 |** Randomly selected, segmented holdout glomeruli.
A batch of randomly selected glomeruli with the predicted boundaries - from the 100 holdout WSIs. This selection is intended to highlight the diversity pathology and staining of the holdout dataset.